\documentclass[preprint]{emulateapj}

\shorttitle{Discs around Blue Stragglers?}
\shortauthors{Porter \& Townsend}

\newcommand{\Msun}{\ensuremath{M_{\sun}}}
\newcommand{\Rsun}{\ensuremath{R_{\sun}}}

\newcommand{\Mstar}{\ensuremath{M_{\ast}}}

\newcommand{\Rpole}{\ensuremath{R_{\rm p}}}

\newcommand{\Kelv}{\ensuremath{\rm K}}
\newcommand{\kms}{\ensuremath{\rm km s^{-1}}}

\newcommand{\Teff}{\ensuremath{T_{\rm eff}}}
\newcommand{\Teffz}{\ensuremath{T_{\rm eff, 0}}}
\newcommand{\geff}{\ensuremath{g_{\rm eff}}}

\newcommand{\vsini}{\ensuremath{v \sin i}}
\newcommand{\veq}{\ensuremath{v_{\rm eq}}}
\newcommand{\vcrit}{\ensuremath{v_{\rm crit}}}

\newcommand{\Hmin}{\ensuremath{H^{-}}}

\newcommand{\jump}{\ensuremath{\Delta_{\rm B}}}

\newcommand{\eg}{e.g.}

\begin{document}

\submitted{Recieved: 7th October, 2004, Accepted: 16th March, 2005}
\journalinfo{Accepted for Ap.J.Lett.}

\title{On the evidence for discs around Blue Straggler stars}
\author{John M. Porter}
\affil{Astrophysics Research Institute, Liverpool John Moores
University, Twelve Quays House, Egerton Wharf, Birkenhead, CH41
1LD, UK}
\email{jmp@astro.livjm.ac.uk}
\and 

\author{R. H. D. Townsend}
\affil{Bartol Research Institute, University of Delaware, Newark, DE
19716, USA\\
Department of Physics \& Astronomy, University College London, Gower
Street, London, WC1E 6BT, UK}
\email{rhdt@bartol.udel.edu}

\begin{abstract}
Recent observations of blue stragglers by \citet{DeM2004} have
revealed continuum deficits on the blue side of the Balmer
discontinuity, leading these authors to infer the presence of discs
around the stars. This intriguing possibility may throw light on
aspects of the mechanisms responsible for at least some of these
objects; current theories of blue straggler formation invoke stellar
collisions or interacting binaries, both of which appear capable of
forming a circumstellar disc.

However, by synthesizing photospheric spectra for models of rotating
blue stragglers, we demonstrate that the Balmer jump enhancements can
be wholly attributed to the influence of oblateness and gravity
darkening on the formation of the continuum. 
Therefore, we are led to
conclude that the observations of De Marco et al. can be ascribed a
more prosaic explanation, that of rapid stellar rotation arising from
the merger/interaction formation process.

\end{abstract}


\section{Introduction}

Blue stragglers (BSs) are cluster stars having anomalous evolutionary
histories. They are blue stars of intermediate mass
\citep[a few \Msun,][]{Sha1997}, with typical surface temperatures of $\sim
6,000$--$10,000\,\Kelv$ \citep[\eg,][]{Den1999}. Application of
single-star evolution theory indicates an age too young to be
explained by the age of their parent cluster \citep[see the review
by][and references therein]{Liv1993}. This apparent paradox has been
explained by two production paths for BSs: (i) collisions between two
lower-mass stars \citep[\eg\ see the simulations by][]{Sil2001}, and (ii) mass transfer in
moderately wide binary stars \citep[\eg,][]{Bel2002}. Both of these
processes are likely to cause the remnant BS to spin up. Population
studies of BSs have shown that both of these routes may be necessary
to account for the observations \citep{Dav2004}.

Recently, \citet[][hereinafter DM04]{DeM2004} have presented
Hubble Space Telescope (\textit{HST}) observations of BS spectra, and
have argued that they 
detect the signature of a circumstellar disc around some of their
targets. Disc formation is a common theme in BS generation and hence
the DM04 observations could shed light on the different evolutionary
processes responsible for these objects. The evidence advanced for the
presence of discs rests largely on an apparent continuum deficit (by
$\sim 5$--$10$\%), falling on the short wavelength side of the Balmer
discontinuity; DM04 interpret this deficit -- and the corresponding
\emph{enhancement} in the magnitude of the Balmer jump -- as arising
from circumstellar absorption by the presumed disc.

On account of their putative formation mechanisms, BS are expected to
exhibit moderate to high rotation rates. In this letter we investigate
the effect of such rapid rotation on the photospheric continuum, by
developing a spectral synthesis model that correctly accounts for the
oblateness and gravity darkening arising from the centrifugal
force. We describe the model in \S\ref{sec:model}, and use it in
\S\ref{sec:balmer} to explore the evolution of the Balmer jump as the
rotation rate is varied. In \S\ref{sec:obs} we then apply the model to
the blue straggler M3-17, demonstrating that it can successfully reproduce the
observations by DM04. We discuss and summarize our findings in
\S\ref{sec:discussion}.


\section{The Spectral Synthesis Model} \label{sec:model}

In a non-rotating, spherical star, the atmosphere across the entire
surface may be characterized by a single value each for the effective
temperature \Teff\ and gravity $g$. With the introduction of rotation,
however, the outward pull of the centrifugal force distorts the star
into an oblate spheroid. Across the surface of this spheroid, the
effective gravity \geff\ (composed of the vector sum of the Newtonian
gravity and the centrifugal acceleration) is non-uniform, decreasing
toward the equatorial regions where the centrifugal force is
strongest. Likewise, in accordance with the gravity darkening law of
\citet{vonZ1924}, the effective temperature decreases toward the
equator in accordance with the relation $\Teff \propto \geff^{\case{1}{4}}$.

To model such an oblate, gravity-darkened star, we set up a raster
grid that divides the surface\footnote{Defined by an isosurface of the
effective potential within the Roche approximation \citep[see,
\eg,][]{Cra1996}.} into c. 14,000 pixels\footnote{The exact number of
pixels composing the surface varies with the degree of centrifugal
distortion of the star.}, each covering an area $0.015\,\Rpole$
square, where \Rpole\ is the star's polar radius. We calculate the
local effective gravity associated with the surface point to which
each pixel corresponds, and we likewise assign a local effective
temperature
\begin{equation}
T_{\rm eff} = \left( \frac{4\pi\, \Rpole^{2}\, \geff}
{\Sigma_{1}}\right)^{\case{1}{4}}\, \Teffz,
\end{equation}
following \citeauthor{vonZ1924}'s law. The overall normalization of
these temperature data is specified by the notional effective
temperature \Teffz\ that the star would have if it were non-rotating,
under the (reasonably-accurate) ansatz that the stellar bolometric
luminosity remains invariant as the rotation rate changes. The symbol
$\Sigma_{1}$ denotes the surface-area weighted gravity of the
distorted star, which depends amongst other things on the rotation
rate \citep[see equations 4.22--4.24 of][]{Cra1996}.

Knowing the effective temperature and gravity of each pixel, and
furthermore the local projection cosine $\mu$ of the surface normal
onto the line of sight, we interpolate the observer-directed emergent
flux in a \Teff--$\log g$--$\mu$ grid of precomputed angle-dependent
intensity spectra. By co-adding the flux data from all pixels,
weighted by their projected area $(0.015\Rpole)^{2}$ and
Doppler shifted by the line-of-sight velocity due to rotation, we
thereby build up a disc-integrated spectrum for the entire distorted,
gravity-darkened star. For the calculations presented in the following
section, we adopt an intensity spectrum grid calculated using the
\textsc{synspec} spectral synthesis code of I. Hubeny and T. Lanz. The
spectra incorporate lines due to H, He, C, N, O, Si, Mg, and Ne, and
are based on the am20ak2-odfnew grid of line-blanketed LTE model atmospheres
published by \citet{Kur1993} (these have H/He = 0.34 by mass and are
alpha enhanced by 0.4dex and metal depleted by 2.0dex).

\begin{figure}
\plotone{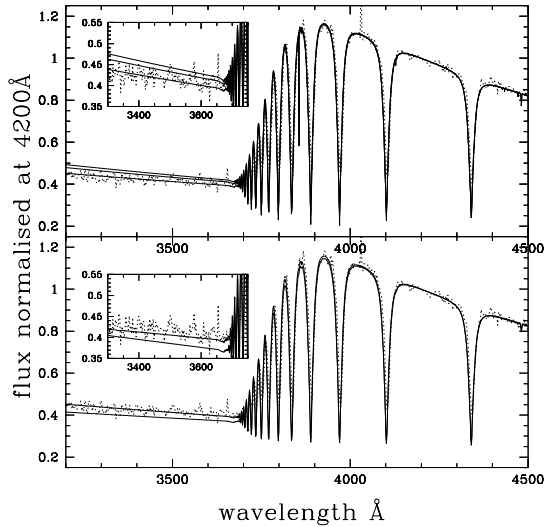}
\caption{Synthetic spectra from rotating stars: the top panel shows a
non-rotating model, and (decreasing downward for wavelengths less than
3700\,\AA) models for $\veq/\vcrit = 0.3$ and 0.5, while the lower
panel have models for $\veq/\vcrit = 0.7$ and 0.9 (increasing
upward). All models are equator-on (inclination $i=90\degr$). The
insets are the models presented over a smaller wavelength range to
highlight the flux shortward of the Balmer Jump. The dotted line on
all panels shows the spectrum of M3-17.}\label{fig:spectra}
\end{figure}

\begin{figure}
\plotone{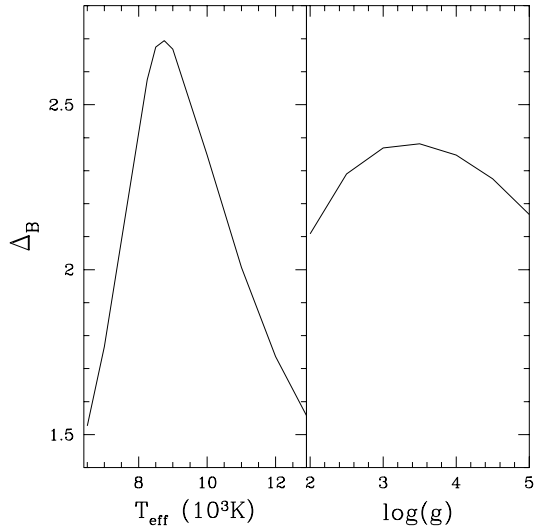}
\caption{The Balmer jump magnitude \jump\ plotted as a function of
effective temperature at fixed $\log g = 4.0$ (left), and as a function
of gravity at fixed $\Teff=10,000\,\Kelv$ (right).}
\label{fig:balmer-t-g}
\end{figure}

\section{Influence of Rotation on the Balmer Jump and Colors} \label{sec:balmer}In Fig.~\ref{fig:spectra} we present synthetic spectra covering the
wavelength range 3,200--4,500\,\AA, calculated for equatorial rotation
rates $\veq/\vcrit = 0.0, 0.3, 0.5, 0.7$ and 0.9 using the model we
describe above; here, $\vcrit \equiv \sqrt{2 G \Mstar/3 \Rpole}$ is
the critical rotation velocity at which the equatorial centrifugal
force balances gravity. The underlying star has a mass $\Mstar =
1.35\,\Msun$, polar radius $\Rpole = 2.4\,\Rsun$ and `non-rotating'
effective temperature $\Teffz = 10,000\,\Kelv$, these parameters being
chosen to coincide with those given by DM04 for the star M3-17. The
model star is viewed equator-on, and we normalize all spectra to have
a unit flux at 4,200\,\AA.

It is clear that the synthetic spectra are all similar longward
of the Balmer discontinuity at 3,647\,\AA\ (see below) but that the
magnitude of the jump 
-- which we characterize throughout via the ratio $\jump \equiv
f_{4,200}/f_{3,630}$ between the continuum fluxes at 4,200\,\AA\ and
3,630\,\AA\footnote{These points being chosen as well separated from
spectral lines} -- varies noticeably with the rotation rate. In
interpreting this behaviour, we recall that the spectrum of a rotating
star is a composite, made up from contributions covering a range of
effective temperatures and gravities. To illustrate the sensitivity of
the Balmer jump against such variation in temperature and gravity,
Fig.~\ref{fig:balmer-t-g} plots \jump\ as a function of both \Teff\
and $\log g$, for a spectrum synthesized from a non-rotating
plane-parallel atmosphere model. 

The temperature dependence of \jump\ (left-hand panel) exhibits a
sharp peak around $\Teff \sim 8,500\,\Kelv$; at temperatures cooler
than this turnover, the appearance of \Hmin\ bound-free opacity --
which preferentially absorbs continuum photons toward longer
wavelengths in the optical and UV \citep[see, \eg\
Fig.~8.3 of][]{Gra1992} -- tends to suppress the flux redward of the Balmer
discontinuity, resulting in a reduction of \jump. Likewise, at
temperatures hotter than $8,500\,\Kelv$, the progressive depopulation
of the $n=2$ (Balmer ground-state) level of neutral hydrogen removes
bound-free continuum opacity blueward of the discontinuity, again
resulting in a reduction of \jump.

Similar behaviour is exhibited in the gravity dependence of the Balmer
jump (right-hand panel of Fig.~\ref{fig:balmer-t-g}); \Hmin\ number
densities are enhanced at high 
gravities, due to the corresponding increase in electron density,
while $n=2$ level populations are depleted at low gravities, due to
the reduction in collisional recombinations which replenish these
populations. Together, these processes are responsible for the decline
in \jump\ toward both low and high $g$.

\begin{figure}
\plotone{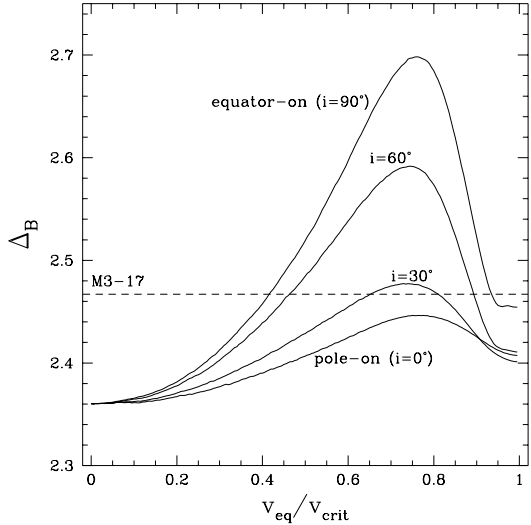}
\caption{The Balmer jump magnitude \jump\ plotted as a function of
rotation rate $\veq/\vcrit$, for the stellar model representative of
M3-17 (see text). Data for different inclinations are shown;
the horizontal dashed line indicates the jump magnitude observed in
M3-17.}
\label{fig:balmer-rot}
\end{figure}

In Fig.~\ref{fig:balmer-rot}, we illustrate how the combined
\Teff/$\log g$ sensitivities of the Balmer-jump magnitude come
together in a rotating star, by plotting \jump\ as a function of
rotation rate. We use the same stellar parameters as previously, but
in the present case show data for both pole- and equator-on
aspects. In each, \jump\ increases steadily up to a rate $\veq/\vcrit
\sim 0.77$, due to the reduction in equatorial effective temperature
and gravity from their non-rotating values $\Teff=10,000\,\Kelv$,
$\log \geff = 3.81$. At this point, the equatorial reduction of
$\Teff$ and \geff\ is so pronounced that the low temperature/gravity
regime of \jump\ is reached, and the magnitude of the Balmer jump then
decreases rapidly toward even-higher \veq. However, in every case,
\jump\ for the rotating models is larger than in the non-rotating
limit.

\begin{figure}
\plotone{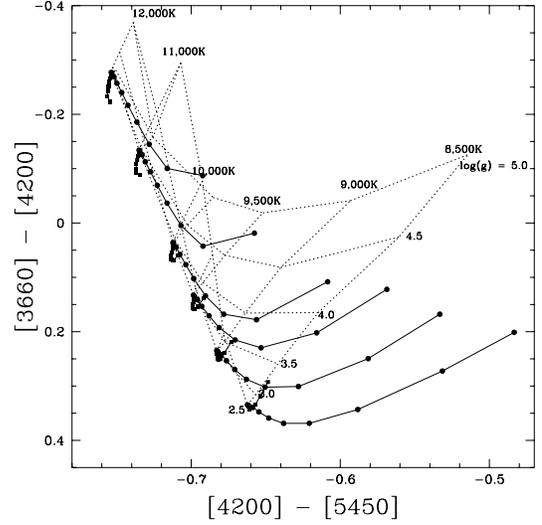}
\caption{Color-color figure for both rotating and non-rotating
  atmosphere models. The dotted lines (with effective temperatures and
  gravities marked) correspond to models with zero rotation. The solid
  lines correspond to models with rotation rates of $v/v_{\rm crit}$ =
0.0--0.9 in steps of 0.1 for pole-on (squares) and
  equator-on (circles) models.}
\label{fig:color-color}
\end{figure}

How then does rotation change the flux distribution redward of the
Balmer Jump?  
In Fig.~\ref{fig:color-color}, we present a color-color diagram of
the rotating models. The fluxes in the three photometric bands 
(centered at 3660\AA, 4200\AA\ and 5450\AA) are averaged over 120\AA,
80\AA\ and 120\AA\ widths 
respectively, and we use the zero-point magnitudes from 
\citet{Bessell1998} -- this is exactly the same as the procedure
used DM04 to allow a direct comparison to be made.
A grid of non-rotating model atmospheres is plotted (dotted lines) for
temperatures between 8,500K--12,000K, and for log$g$=2.5-5.0. 
Also we have calculated 
pole-on and equator-on rotating atmosphere models (for
$v_{\rm eq}/v_{\rm crit}$ = 0.0--0.9 in steps of 0.1) for the [4200]-[5450]
extreme blue edge of the non-rotating grid.
The equator-on models (filled circles) sweep down and to the
right (almost appearing like models becoming cooler at constant log$g$),
with the lower-temperature models eventually sweeping upward in a broad ``u''
shape. The loci of the pole-on models (filled squares) 
initially move vertically down with increasing rotation (i.e. no change in the
[4200]-[5450]) with the cooler models then reversing their
track in a ``v'' shape. The maximum change in [3660]-[4200] becomes
more pronounced for higher temperature models, reaching 0.06mag for the
12,000K model.

Two conclusions can be drawn from this figure: first, that for pole-on
models, rotation can cause the Balmer jump (signified by the
[3660]-[4200] color) to be too large whilst keeping the [4200]-[5450]
color constant; second, if they are rotating, stars can exist with
colors which are apparently inconsistent with non-rotating model
atmospheres (i.e. the lower left-hand extremum of the non-rotating
atmosphere grid is extended a little). Finally we note that the
position of a star in the 
color-color plane does not yield a unique pair of effective
temperature and gravity values -- two color measurements cannot
produce three unique values of effective temperature, gravity, and
rotation. 

\section{Application to M3-17} \label{sec:obs}

The observed projected rotation velocity of M3-17, $\vsini = 200\pm
50\,\kms$, along with the stellar parameters furnished by DM04, implies that
this object is a rapid rotator, having $\veq\sin i/\vcrit = 0.75\pm
0.18$. The Balmer jump magnitude derived from the \textit{HST}
spectrum of this object is $\jump = 2.47$, a value that we indicate in
Fig.~\ref{fig:balmer-rot} by the horizontal dashed line. Clearly,
there are multiple rotating models that can fit the observations,
without the need to invoke a circumstellar disc; for example, a match
to \jump\ can be achieved with an edge-on ($i=90\degr$) model rotating
at $\veq/\vcrit \approx 0.4$ or 0.9, and other solutions can be found
at lesser inclinations.

We further illustrate this point in Fig.~\ref{fig:spectra}, where we
plot the archival spectrum of M3-17 over our model data. The
non-rotating synthetic spectrum clearly has a surfeit of continuum
flux shortward of the Balmer discontinuity, which led DM04 to diagnose
the presence of a circumstellar disc. However, our equator-on rotating
models at $\veq/\vcrit = 0.4$ and 0.9 clearly provide a good fit to
the continuum flux, suggesting that rotation alone may suffice to
explain the Balmer jump.
In support of this conclusion, we note that the rotational
reddening of the Paschen continuum, as seen for the equator-on curves
in Fig.~\ref{fig:color-color}, actually enables the rotating models to
fit the observed [4200]-[5450] color of M3-17 \emph{better} than the
non-rotating model of the same temperature.
We do observe that our models are unable to reproduce the strengths of the
Balmer-series absorption lines seen redward of 3,647\AA; this problem
reflects the fact that we specify the model temperature \emph{a
priori} from that of DM04. The issue can be resolved by allowing
\Teffz\ (and other parameters) to vary under the control of a
suitably-chosen fitting statistic.
However, we do not undertake such fine tuning here, because we
are not attempting to derive exact atmospheric parameters for M3-17
here. Rather, our objective is to highlight that high rotation has a
significant effect on the continuum, and may mimic the presence of an
obscuring disk.

\section{Discussion \& Summary} \label{sec:discussion}

Using a spectral synthesis approach that correctly treats oblateness
and gravity darkening, we have demonstrated that apparent flux
deficits shortward of the Balmer discontinuity can be attributed
wholly to rapid rotation. This casts doubt on the disc hypothesis
advanced by \citet{DeM2004} to explain the Balmer jump anomalies seen
in the spectra of BSs. At least in the case of rapid rotator M3-17
($\vsini = 200\pm 50\,\kms$), the continuum is well fit by our model,
leaving little reason to invoke the presence of a disc.

In addition to M3-17, five other stars (out of a total of 50) in the
DM04 sample show Balmer jumps that are enhanced with respect to
non-rotating models\footnote{By this, we mean models that do not
include oblateness or gravity darkening.}. Of these five, only
NGC6751-11 has a measured (upper limit) projected rotation velocity, $\vsini <
50\,\kms$. Due to the unknown projection factor $\sin i$, the upper
limit on the intrinsic equatorial velocity \veq\ may be much larger,
and it is entirely possible that, along with the remaining objects
with anomalous Balmer jumps, this star is a rapid rotator. 

Accordingly, Balmer jump anomalies seen in BSs may \emph{all} be
attributable to rapid rotation. This hypothesis is lent some support
by the fact that the proposed formation mechanisms for BSs --
collisions or mass transfer -- both involve the deposition of
significant amounts of angular momentum on the star, thereby
spinning it up \citep[see, \eg,][]{PolMar1994}.

On a final note, we stress that we do not (and cannot) rule out the
possibility of discs around BSs. Indeed, observations of emission
lines in these objects, bearing a close resemblance to those seen in
Be stars \citep[see][]{Mer1982}, indicate that there almost certainly
\emph{are} circumstellar discs around some of them. However, the
present paper demonstrated why Balmer jump anomalies cannot be used
as a clear and unambiguous diagnostic for BS discs.


\acknowledgments 
RHDT acknowledges support from PPARC and from NSF grant AST-0097983.
We thank the referee for their useful comments on the submitted paper.


\begin{thebibliography}{}
\bibitem[{{Bellazzini} {et~al.}(2002){Bellazzini}, {Fusi Pecci}, {Messineo},
  {Monaco}, \& {Rood}}]{Bel2002}
{Bellazzini}, M., {Fusi Pecci}, F., {Messineo}, M., {Monaco}, L., \&  {Rood},
  R.~T. 2002, \aj, 123, 1509
\bibitem[{{Bessell Castelli \& Plez}(1998)}]{Bessell1998}
{Bessell}, M.S., {Castelli}, F., \& {Plez} B. 1998, \aap, 333, 231
\bibitem[{{Cranmer}(1996)}]{Cra1996}
{Cranmer}, S.~R. 1996, PhD thesis, University of Delaware
\bibitem[{{Davies} {et~al.}(2004){Davies}, {Piotto}, \& {de Angeli}}]{Dav2004}
{Davies}, M.~B., {Piotto}, G., \&  {de Angeli}, F. 2004, \mnras, 349, 129
\bibitem[{{De Marco} {et~al.}(2004){De Marco}, {Lanz}, {Ouellette}, {Zurek}, \&
  {Shara}}]{DeM2004}
{De Marco}, O., {Lanz}, T., {Ouellette}, J.~A., {Zurek}, D., \&  {Shara}, M.~M.
  2004, \apj, 606, L151
\bibitem[{{Deng} {et~al.}(1999){Deng}, {Chen}, {Liu}, \& {Chen}}]{Den1999}
{Deng}, L., {Chen}, R., {Liu}, X.~S., \&  {Chen}, J.~S. 1999, \apj, 524, 824
\bibitem[{{Gray}(1992)}]{Gra1992}
{Gray}, D.~F. 1992, {The Observation and Analysis of Stellar Photospheres}, 2nd
  edn. Cambridge: Cambridge University Press
\bibitem[{{Kurucz}(1993)}]{Kur1993}
{Kurucz}, R. 1993, CD-ROM No. 16. Smithsonian Astrophysical Observatory,
  Washington D.C.
\bibitem[{{Livio}(1993)}]{Liv1993}
{Livio}, M. 1993, in ASP Conf. Ser. 53, Blue Stragglers, ed. R.~E. {Staffer},
  San Francisco: ASP, 3
\bibitem[{{Mermilliod}(1982)}]{Mer1982}
{Mermilliod}, J.-C. 1982, \aap, 109, 37
\bibitem[{{Pols} \& {Marinus}(1994)}]{PolMar1994}
{Pols}, O.~R., \&  {Marinus}, M. 1994, \aap, 288, 475
\bibitem[{{Shara} {et~al.}(1997){Shara}, {Saffer}, \& {Livio}}]{Sha1997}
{Shara}, M.~M., {Saffer}, R.~A., \&  {Livio}, M. 1997, \apj, 489, L59
\bibitem[{{Sills} {et~al.}(2001){Sills}, {Faber}, {Lombardi}, {Rasio}, \&
  {Warren}}]{Sil2001}
{Sills}, A., {Faber}, J.~A., {Lombardi}, J.~C., {Rasio}, F.~A., \&  {Warren},
  A.~R. 2001, \apj, 548, 323
\bibitem[{{von Zeipel}(1924)}]{vonZ1924}
{von Zeipel}, H. 1924, \mnras, 84, 665
\end{thebibliography}
\end{document}